\documentclass[epj-spec]{svjour}
\usepackage{graphics}
\begin{document}
\title{Highlights and Perspectives of the JLab Spin Physics Program}
\author{Jian-ping Chen\inst{1}\fnmsep{\email{jpchen@jlab.org}}}
\institute{Thomas Jefferson National Accelerator Facility, Newport News, 
VA 23606, USA}
\abstract{
Nucleon spin structure has been an active, exciting and intriguing subject 
of interest for the last three decades.
Recent precision spin-structure data from Jefferson Lab have significantly 
advanced our knowledge of nucleon structure in the valence quark
(high-$x$) region and improved our understanding of
higher-twist effects, spin sum rules and 
quark-hadron duality. 
First, results of spin sum rules and polarizabilities in the low to 
intermediate $Q^2$ region are presented. Comparison with theoretical
calculations, in particular with Chiral Perturbation Theory (ChPT) 
calculations, are discussed. Surprising disagreements of ChPT calculations
with experimental results on the generalized spin polarizability, 
$\delta_{LT}$, were found. 
Then, precision measurements of the spin 
asymmetry, $A_1$, in the high-$x$ region are presented.
They provide crucial input for global fits to world data to extract 
polarized parton distribution functions. 
The up and down quark spin distributions in the nucleon were extracted. 
The results for $\Delta d/d$ disagree with the leading-order pQCD prediction 
assuming hadron helicity conservation.
Results of precision measurements of 
the $g_2$ structure function to study higher-twist effects are presented.
The data indicate a significant higher-twist (twist-3 or higher) effect.
The second moment of the spin structure functions and the twist-3
matrix element $d_2$ results were 
extracted. The high $Q^2$ result was compared
with a Lattice QCD calculation.
Results on the resonance spin-structure functions in the intermediate
$Q^2$ range are presented, which, in combination with DIS
data, enable a detailed study of quark-hadron duality in spin-structure 
functions.
Finally, an experiment to study neutron transversity and transverse 
spin asymmetries is discussed. 
A future plan with the 
12 GeV energy upgrade at JLab is briefly outlined.
} 
\maketitle
\section{Introduction}
\label{intro}
In the last three decades the study of the spin structure of the nucleon has
led to a very productive
experimental and theoretical activity with exciting results and new challenges\cite{spin}. This
investigation has included a variety of aspects, such as
testing QCD in its perturbative regime 
{\emph {via}} spin sum rules (like the Bjorken sum rule\cite{Bjorken}) and understanding how the spin of the nucleon is
built from the intrinsic degrees of freedom of the theory, quarks and gluons. 

Recently, the high polarized luminosity available at Jefferson 
Lab has allowed the study of nucleon spin structure with 
an unprecedented precision, enabling us to access the
valence quark (high-$x$) region\cite{e99117} and also to expand the study to the
second spin-structure function, $g_2$~\cite{e97103}. Furthermore, the moments of
the spin-structure functions~\cite{chen05} were measured~\cite{e94010} and the spin sum 
rules~\cite{e94010,bjsum}, polarizabilities~\cite{e94010}
and quark-hadron duality~\cite{e01012,RSS} were studied. 
A new program to study the transverse spin and transverse momentum dependence 
in the nucleon is under way.

\subsection{Inclusive polarized electron-nucleon scattering}
\par
For inclusive polarized electron scattering off a polarized 
nucleon target, 
the cross section depends on four structure functions, $F_1(Q^2,x)$, 
$F_2(Q^2,x)$, $g_1(Q^2,x)$ and $g_2(Q^2,x)$, where 
$F_1$ and $F_2$ are the unpolarized structure functions 
and $g_1$ and g$_2$ the polarized structure functions. 
In the quark-parton model, 
$F_1$ or $F_2$ gives the quark momentum 
distribution and $g_1$ gives the quark spin distribution.
Another physics quantity of interest is the virtual photon-nucleon 
asymmetry $A_1$
\begin{equation}
A_1={g_1-(Q^2/\nu^2) g_2 \over F_1} \approx {g_1 \over F_1}.
\end{equation}

\subsection{Moments and sum rules of the spin-structure functions}
\par
Sum rules involving the spin structure of the nucleon offer an important opportunity to study QCD. In recent years
the Bjorken sum rule at large $Q^2$ and 
the Gerasimov, Drell and Hearn (GDH) sum rule~\cite{gdh} at $Q^2=0$
have attracted large experimental and theoretical~\cite{dre01} efforts that have provided us with rich information. 
A generalized GDH sum rule~\cite{ggdh} connects the GDH sum rule with the 
Bjorken sum 
rule and provides a clean way to test theories with experimental data over the
entire $Q^2$ range. 
Spin sum rules relate the moments of the spin-structure functions to the 
nucleon's static properties (as in Bjorken or GDH sum rules) or
real or virtual Compton amplitudes, which can be calculated theoretically
(as in the generalized GDH sum rule or the forward spin polarizabilities). 
Refs.~\cite{chen05,dre} provide comprehensive reviews on this subject.

\subsection{Spin structure in the valence quark (high-$x$) region}
\par
The high-$x$ region is of special interest, because this is where the valence 
quark contributions are expected to dominate.
With sea quark and explicit gluon contributions expected not to be
important, it is a clean region to test our understanding of nucleon
structure. Relativistic constituent quark models\cite{vqm}
should be applicable in this region
and perturbative QCD\cite{pQCD} can be used to make predictions in the large
$x$ limit. 

To first approximation, the constituent quarks in the nucleon are
described by SU(6) wavefunctions.
SU(6) symmetry leads to the following predictions\cite{SU6}: 

\begin{equation}
A_1^p=5/9;\ \ A_1^n=0; \ \ \Delta u/u=2/3; \ \ 
\Delta d/d=-1/3.
\label{eq:SU6}
\end{equation}

Relativistic Constituent Quark Models (RCQM) with broken SU(6) symmetry, e.g., 
the hyperfine 
interaction model\cite{vqm}, lead to a dominance of a `diquark' 
configuration 
with the diquark spin $S=0$ at high $x$. This implies that as $x\rightarrow1$:
\begin{equation}
 A_1^p\rightarrow 1;\ \
   A_1^n\rightarrow 1;\ \ \Delta u/u \rightarrow 1;\ \ {\rm and} \ \ 
\Delta d/d \rightarrow -1/3.
\label{eq:rnpqcd}
\end{equation}
\noindent In the RCQM, relativistic effects lead to a 
non-zero quark orbital angular momentum and reduce the valence quark 
contributions to the nucleon spin from 1 to $0.6 - 0.75$.
 
Another approach is leading-order pQCD~\cite{pQCD}, which assumes the 
quark orbital angular momentum to be negligible and leads to hadron helicity 
conservation. 
It yields:  

\begin{equation}
A_1^p\rightarrow 1;\ \
   A_1^n\rightarrow 1;\ \ 
\Delta u/u \rightarrow 1;\ \ {\rm and} \ \ 
\Delta d/d \rightarrow 1.
\label{eq:rnppqcd}
\end{equation}
\noindent
Not only are the limiting values as $x\rightarrow 1$ important, but also
the behavior in the high-$x$ region. The behavior of $A_1$  
as $x$ approaches 1 is sensitive to
the dynamics in the valence quark region. 

\subsection{The $g_2$ structure function and the $d_2$ moment}
\par
The spin structure function $g_2$, unlike $g_1$ and $F_1$, can not be
interpreted in the simple quark-parton model. To understand $g_2$ properly, 
it is best to start with the operator product expansion (OPE) method.
In the OPE, neglecting quark masses, $g_2$ can be cleanly separated into a
twist-2 and a higher twist term:
  \begin{eqnarray}g_2(x,Q^2) = g_2^{WW}(x,Q^2) +g_2^{H.T.}(x,Q^2)~.
  \end{eqnarray}
The leading-twist term can be determined from 
$g_1$ as\cite{WW}
  \begin{eqnarray}
   g_2^{WW}(x,Q^2) = -g_1(x,Q^2) + \int _{x}^1 \frac{g_1(y,Q^2)}{y} dy~,
  \end{eqnarray}
and the higher-twist term arises from quark-gluon correlations.
Therefore, $g_2$ provides a clean way to study higher-twist effects.
In addition, at high $Q^2$, the $x^2$-weighted moment, $d_2$, 
is a twist-3 matrix element and is related to the color 
polarizabilities~\cite{d2}:
\begin{equation}
d_2 = \int _{0}^{1} x^2 [g_2(x)-g_2^{WW}(x)] dx.
\end{equation}
Predictions for $d_2$ exist from various models~\cite{str,wei,wak} and 
lattice QCD\cite{LQCD}.

\subsection{Quark-hadron duality in spin structure functions}
\par
Quark-hadron duality was first observed  
for the spin-independent structure function $F_2$. In 1970, Bloom and Gilman~\cite{BG} noted that the nucleon 
resonance data averaged to the DIS scaling curve. Recent precision 
data~\cite{F2dual} confirm quark-hadron duality in the unpolarized proton
structure functions. Efforts are ongoing to investigate quark-hadron duality
in polarized structure functions~\cite{MEK}. It was predicted that in 
the high-$x$ region at high enough $Q^2$, the resonances will have a similar 
behavior as DIS. Results from HERMES~\cite{HERMES} and CLAS~\cite{eg1b} 
show the proton spin structure function $g_1^p$ approaching
duality. The study of 
quark-hadron duality will aid in the study of the higher-twist effects and 
understanding the 
high-$x$ behavior in DIS. 

\subsection{Transversity}
\par
The transversity distributions, $\delta q(x,Q^2)$,
are fundamental leading-twist (twist-2) quark distributions, samilar to 
the unpolarized
and polarized parton distributions, $q(x,Q^2)$ and $\Delta q(x,Q^2)$. 
In quark-parton models,
they describe the net transverse
polarization of quarks in a transversely polarized nucleon.
There are several special features for the transversity distributions,
making them uniquely interesting: 

\begin{itemize}
    \item
The difference between the transversity and the longitudinal distributions
is purely due to relativistic effects. In the absence of relativistic
effects (as in the non-relativistic quark model, where boosts and rotations
commute), the transversity distributions are identical
to the longitudinally polarized distributions.

    \item
          The quark transversity distributions do not mix with gluonic
effects~\cite{BST} and therefore follow a much simpler evolution and have
          a valence-like behavior.

    \item The positivity of helicity amplitudes leads to the 
Soffer's inequality for the transversity\cite{soffer}:
    $\vert h_1^q \vert \le \frac {1}{2} (f_1^q + g_{1}^q)$.

    \item The lowest moment of $h_1^q$ measures a simple local
    operator analogous to the axial charge, known as the ``tensor charge'',
which can be calculated from lattice QCD.
\end{itemize}

Due to the chiral-odd nature of the transversity distribution, it can not
be measured in inclusive DIS experiments. In order 
to measure $\delta q(x,Q^2)$,
an additional chiral-odd object is required, such as double-spin asymmetries in
Drell-Yan processes, single target-spin azimuthal asymmetries in
Semi-Inclusive DIS (SIDIS) reactions, double-spin asymmetries in 
$\Lambda$ production
from e-p and p-p reactions and single-spin asymmetries in double pion 
production from e-p scattering.
The first results, from measurements performed by the HERMES~\cite{hermes03} and COMPASS~\cite{compass} collaborations with SIDIS offered the first glimpse of possible effects caused by the transversity distributions.

\section{Recent results from Jefferson Lab}
\par
The Thomas Jefferson National Accelerator Facility (Jefferson Lab, or JLab, 
formerly known as CEBAF - Continuous Electron Beam Accelerator Facility)
is located in Newport News, Virginia, USA. The accelerator produces 
a continuous-wave electron beam
of energy up to 6 GeV. An energy upgrade to 12 GeV is planned in the next
few years. The electron beam with a current of up to 180 $\mu$A is 
polarized up to $85\%$ by illuminating a 
strained GaAs cathode with polarized laser light. The electron beam is deflected 
to three experimental halls (Halls A, B and C) where the electron beam can be
scattered
off various nuclear targets. The scattered electrons and knocked 
out particles are detected in the halls with various spectrometer detector 
packages. The experiments reported here are from inclusive electron scattering
where only the scattered electrons are detected.
The neutron results presented here are from
Hall A~\cite{NIMA} where there are two High Resolution Spectrometers (HRS) 
with momentum up to 4 GeV/$c$. A polarized $^3$He target, with in-beam polarization of about $50\%$,
provides an effective polarized neutron target. The polarized luminosity reached is $10^{36}$ s$^{-1}$cm$^{-2}$.
The detector package consists of vertical
drift chambers (for momentum analysis 
and vertex reconstruction), scintillation counters (data acquisition 
trigger) and \v{C}erenkov counters and lead-glass calorimeters 
(for particle identification (PID)). The $\pi^-$ were sorted from e$^-$ with an
efficiency better than 99.9\% .
Both HRS spectrometers were used to double the statistics and constrain the 
systematic uncertainties by comparing the cross sections extracted using each HRS.
The proton and deuteron results are from 
Hall B~\cite{NIMB}, where there is the CEBAF Large Acceptance Spectrometer (CLAS) 
and Hall C, where there are the High Momentum Spectrometer (HMS) and the Short Orbit Spectrometer (SOS). Polarized solid $NH_3$ and $ND_3$ 
targets~\cite{NH3} using dynamical nuclear polarization were used. Polarizations
up to $95\%$ for $NH_3$ and up to $45\%$ for $ND_3$ were achieved.

\subsection{Results of the generalized GDH sum and BC sum for $^3$He and the neutron}
\label{GDHsum}
\begin{figure}[!ht]
\parbox[t]{2in}{\scalebox{0.36}{\centering\includegraphics[0,52][422,535]{GDH_He3.eps}}}
\parbox[t]{2in}{\scalebox{0.34}{\centering\rotatebox{-90}{\includegraphics[550,200][872,725]{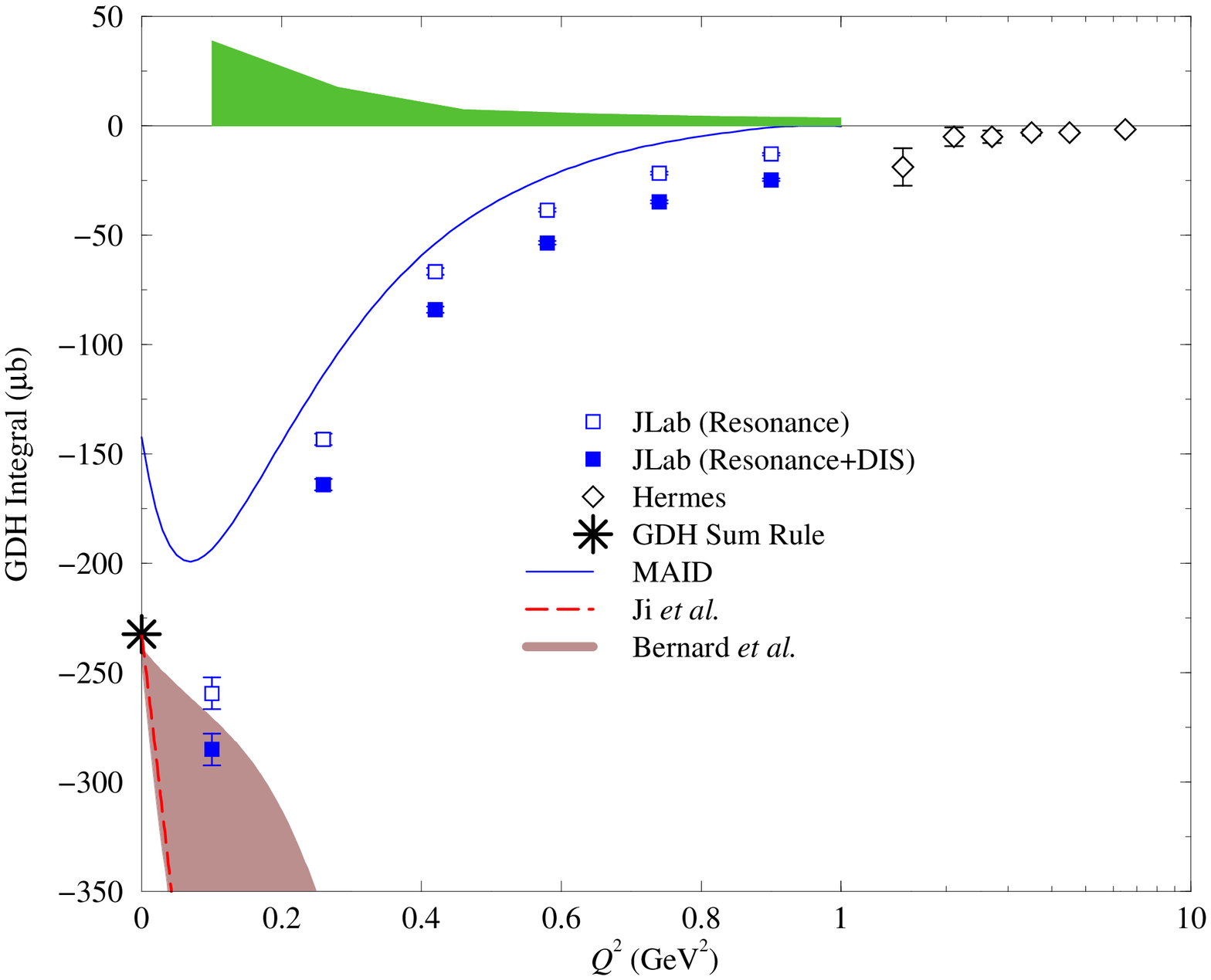}}}}
\parbox[t]{2in}{\scalebox{0.36}{\centering\includegraphics[0,-350][422,233]{BC_He3.eps}}}
\parbox[t]{2in}{\scalebox{0.34}{\centering\rotatebox{-90}{\includegraphics[1020,-150][1442,450]{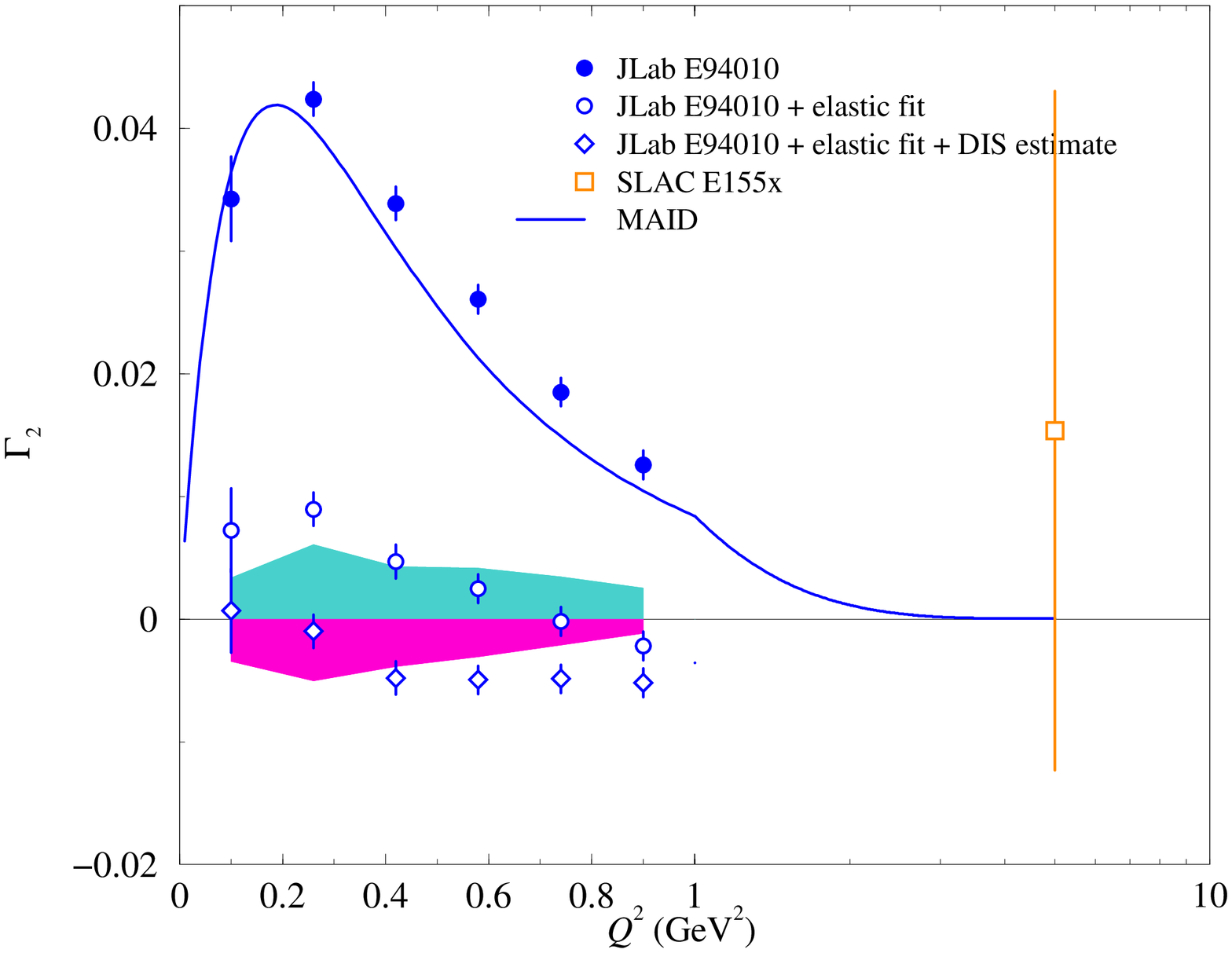}}}}
\vspace {-9.7cm}
\caption{Results of GDH sum $I(Q^2)$ and BC sum $\Gamma_2 (Q^2)$ for $^3$He~\protect\cite{e94010he3}
 and the neutron~\protect\cite{e94010}. The $^3$He GDH results are compared with the MAID model plus quasielastic contribution. The neutron GDH results are compared with $\chi$PT calculations of 
ref.~\protect\cite{chpt} (dotted line) and ref.~\protect\cite{chpt2} 
(dot-dashed line).  
The MAID model calculation of ref.~\protect\cite{dre01},
is represented by a solid line.  Data from HERMES~\protect\cite{HERMES} are also shown. The BC sum results (resonance only) are compared with MAID model calculations.}
\label{fig:GDH}
\end{figure}

Fig.~\ref{fig:GDH} shows the extended GDH integrals  
$I(Q^2)=\int_{\nu_thr}^\infty [\sigma_{1/2}(Q^2)-sigma_{3/2}d\nu/nu(Q^2)]$ 
(open circles) for $^3$He (preliminary) (upper-left) and  for the neutron (upper-right),
which were extracted from Hall A experiment E94-010\cite{e94010}, from break-up threshold for $^3$He (from pion threshold for the neutron) to $W=2$ GeV.
The uncertainties, when visible, represent statistics only; the systematics are shown by the grey band.   
The solid squares include an estimate of the unmeasured high-energy part.
The corresponding
uncertainty is included in the systematic uncertainty band.
The preliminary $^3$He results rise with decreasing $Q^2$. Since the GDH sum 
rule at $Q^2=0$ predicts a large negative value, a drastic turn around 
should happen at $Q^2$ lower than 0.1 GeV$^2$. A simple model using MAID~\cite{dre01} plus quasielastic contributions estimated from a PWIA model~\cite{ciofi2} indeed shows the expected turn
around. The data at low $Q^2$ should be a good test ground for few-body Chiral Perturbation Theory calculations.

The neutron results indicate a smooth variation of $I(Q^2)$  to increasingly negative values as $Q^2$ varies from $0.9\,{\rm GeV^2}$ 
towards zero.
The data are more negative than the MAID model calculation\cite{dre01}. 
Since the calculation only includes contributions to $I(Q^2)$ for $W \le 2\,{\rm GeV}$, it should be compared with the
open circles. The GDH sum rule 
prediction, $I(0)=-232.8\,\mu{\rm b}$, is indicated in Fig.~\ref{fig:GDH}, along with  
extensions to $Q^2>0$ using two next-to-leading order $\chi$PT
calculations, one using the Heavy Baryon approximation (HB$\chi$PT)~\cite{chpt} 
(dotted line) and the other Relativistic Baryon $\chi$PT (RB$\chi$PT)~\cite{chpt2} (dot-dashed line). Shown with a grey band is RB$\chi$PT including resonance effects~\cite{chpt2}, which have an associated
large uncertainty due to the resonance parameters used. 

 The capability to transversely polarize the Hall A $^3$He target 
allows precise measurements of $g_2$. 
The integral of $\Gamma_2^{^3He}=\int_0^\infty g_2^{3He}(Q^2) dx$
and $\Gamma_2^n$ 
is plotted in the lower-left and lower-right panels of 
Fig.~\ref{fig:GDH} in the measured region (solid circles) and open circles show
the results after adding an estimated
DIS contribution for $^3$He (elastic contribution for the neutron). 
The solid squares (open diamonds)
correspond to the results obtained after adding the elastic contributions for $^3$He,
(adding an estimated DIS 
contribution assuming $g_2 = g_2^{WW}$ for the neutron). The MAID estimate agrees with the general trend but is slightly lower than the resonance data. 
The two bands correspond to the experimental systematic
errors and the estimate of the systematic error 
for the low-$x$ extrapolation. The total results are consistent with the BC 
sum rule. 
The SLAC E155x collaboration\cite{SLAC} previously reported 
a neutron result at high $Q^2$ (open square), which is consistent with zero but with a rather 
large error bar. On the other hand, the SLAC proton result was reported to deviate 
from the BC sum rule by 3 standard deviations. 
\subsection{First moments of $g_1$ and the Bjorken sum}
\begin{figure}[ht!]
\scalebox{0.38}{\centering {\includegraphics {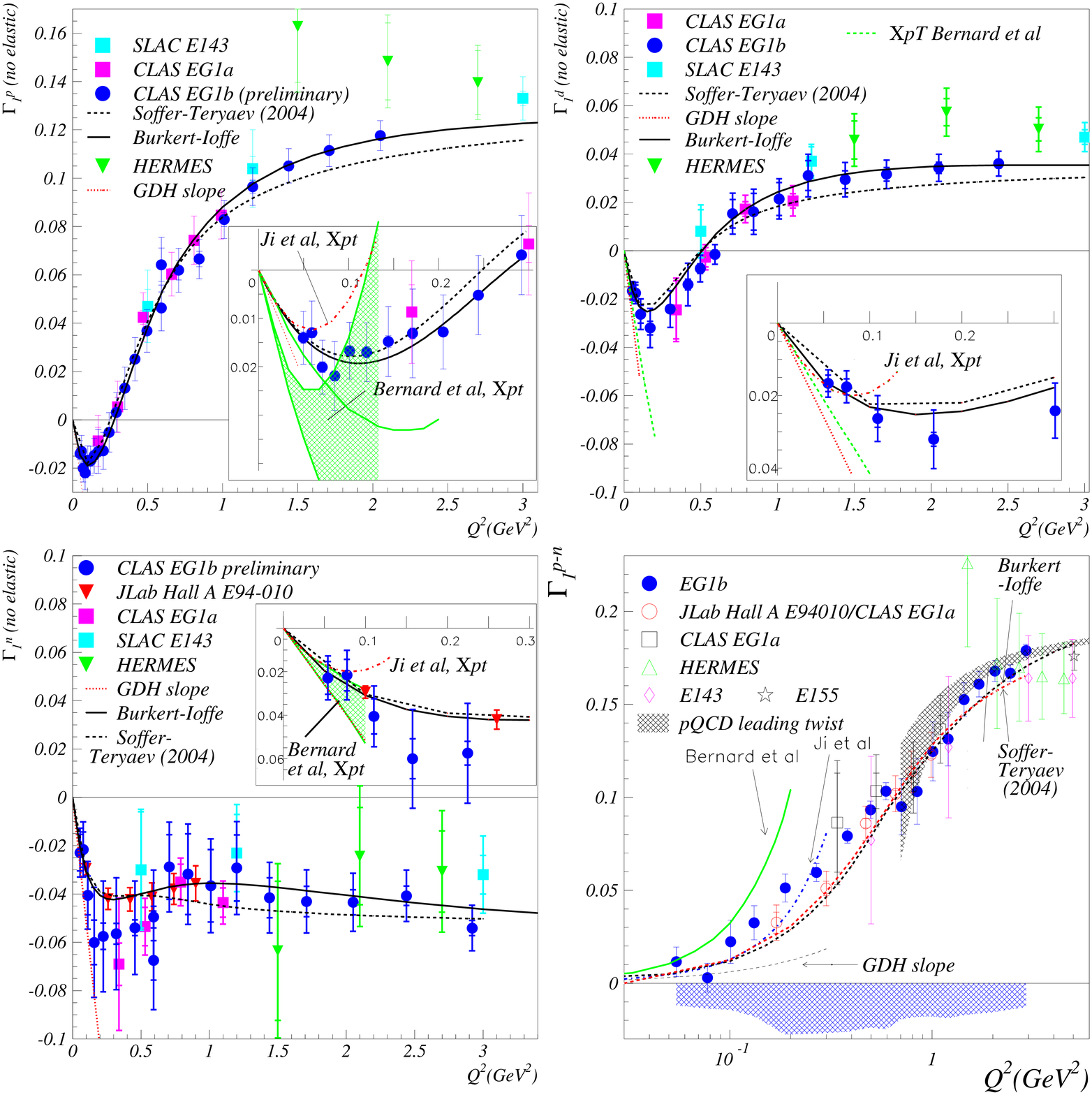}}}
\caption{Results of $\Gamma_1 (Q^2)$ for p, n, d and p-n from JLab Hall A~\protect\cite{e94010} and CLAS eg1a~\protect\cite{eg1a} and eg1b~\protect\cite{eg1b} .
The slopes at $Q^2$=0 predicted by the GDH sum rule 
are given by the dotted lines. The MAID model  
predictions that include only resonance contributions are shown by the full 
lines while 
the dashed (dot-dashed) lines are the predictions from the Soffer-Teryaev (Burkert-Ioffe) model.
The leading twist $Q^2$-evolution of the moments is given by the grey band. 
 The insets show comparisons with $\chi$PT calculations. The full
lines (bands) at low $Q^2$ are the next-to-leading order $\chi$PT 
predictions by Ji \emph{et al.} (Bernard \emph{et al.}). } 
\label{fig:gamma1pn}
\vspace {-4mm}
\end{figure}

The new results from the Hall B eg1b~\cite{eg1b} experiment on $\bar \Gamma_1(Q^2)$ at low to moderate $Q^2$ are shown together with published results from Hall A~\cite{e94010} and Hall B eg1a~\cite{eg1a} in Fig.~\ref{fig:gamma1pn} along with the data from 
SLAC~\cite{SLAC} and HERMES\cite{HERMES}.
The new results are in good agreement with the published data.
 The
inner uncertainty
indicates the statistical uncertainty while the outer one is the quadratic sum of the 
statistical and systematic uncertainties. 

At $Q^2$=0, the GDH sum rule predicts the slopes of $\Gamma_1$ (dotted lines). 
The behavior at low $Q^2$ can be calculated with $\chi$PT. We show results of calculations by Ji {\it et al.}~\cite{chpt} 
using HB$\chi$PT and by Bernard {\it et al.} with and without~\cite{chpt2} the 
inclusion of vector 
mesons and $\Delta$ degrees of freedom. 
The calculations are in reasonable agreements with the data at the lowest $Q^2$
settings of 0.05 - 0.1 GeV$^2$.
At moderate and large $Q^2$ data are compared with two model calculations~\cite{sof02,bur92} . Both models agree well with the data.  
The leading-twist pQCD evolution is shown by the grey band. It tracks the data down to
surprisingly low $Q^2$, which indicates an overall suppression of higher-twist
effects. 

The lower-right panel in Figure 2 shows the moment of $g_1^p-g_1^n$, the 
generalized Bjorken sum. This is of special interest because it contains 
contributions only from the 
flavor non-singlet (or isovector) part. The data at high $Q^2$ value were used
to test the Bjorken sum rule as one of the fundamental tests of QCD.
They were also used to extract a value of strong coupling constant, $\alpha_s$.
The new JLab data at low $Q^2$ provide interesting information in the low energy region, where the strong interaction is truly strong and non-perturbative.
A new attempt~\cite{alphas} was made to extract an effective strong coupling, $\alpha_s^{eff}$ in the low $Q^2$ region (Figure 3). The extracted $\alpha_s^{eff}$ shows a clear trend of weakening $Q^2$-dependence with decreasing $Q^2$. With the GDH sum 
rule as a limit at $Q^2=0$, a model fit through the extracted $\alpha_s^{eff}$ 
show a loss of $Q^2$-dependence as $Q^2$ approaches zero. This is consistent 
with a conformal behavior, which is important for any attempt to apply
AdS/CFT~\cite{ads} for the strong interaction in the low-energy region.

\begin{figure}[ht!]
\centering {\scalebox{0.40}{\includegraphics {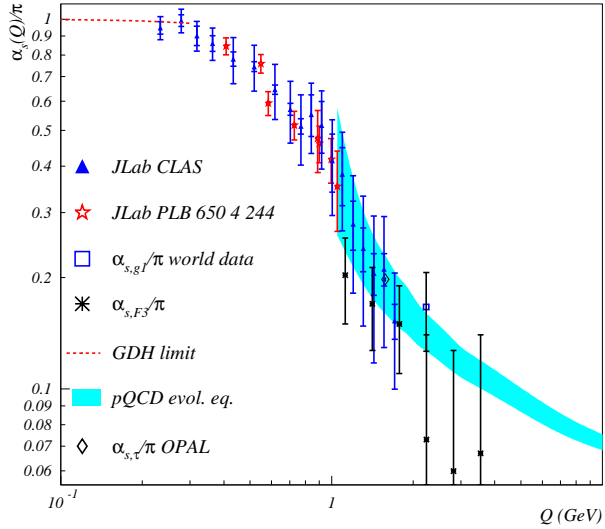}}}
\caption{Results of $\alpha_s^{eff}$~\protect\cite{alphas}
 extracted from generalized Bjorken sum, together with world data on $\alpha_s^{eff}$ and a model fit.}
\end{figure}

\subsection{Spin Polarizabilities: $\gamma_0$ and $\delta_{LT}$}
The generalized spin polarizabilities provide benchmark tests
of $\chi$PT calculations at low $Q^2$.
Since the generalized polarizabilities have an extra $1/\nu^2$ 
weighting compared to the first moments (GDH sum or 
$I_{LT}$), these integrals have less contributions from the large-$\nu$ 
region and converge much faster, which minimizes the uncertainty due to
the unmeasured region at large $\nu$. 

At low $Q^2$, the 
generalized polarizabilities have been evaluated with next-to-leading order 
$\chi$PT 
calculations~\cite{chpt1,chpt2}.
One issue in the $\chi$PT calculations is how to properly
include the nucleon resonance contributions, especially the $\Delta$ resonance.
As was pointed out in Refs.~\cite{chpt1,chpt2} , while $\gamma_0$ is sensitive to 
resonances, $\delta_{LT}$ is insensitive to the $\Delta$ 
resonance. Measurements of the generalized spin
polarizabilities are an important step in understanding the dynamics of
QCD in the chiral perturbation region.

The first results for the neutron generalized forward
spin polarizabilities $\gamma_0(Q^2)$ and $\delta_{LT}(Q^2)$
were obtained at Jefferson Lab Hall A~\cite{e94010}.

\begin{figure}[!ht]
\centering{\scalebox{0.5}{\includegraphics{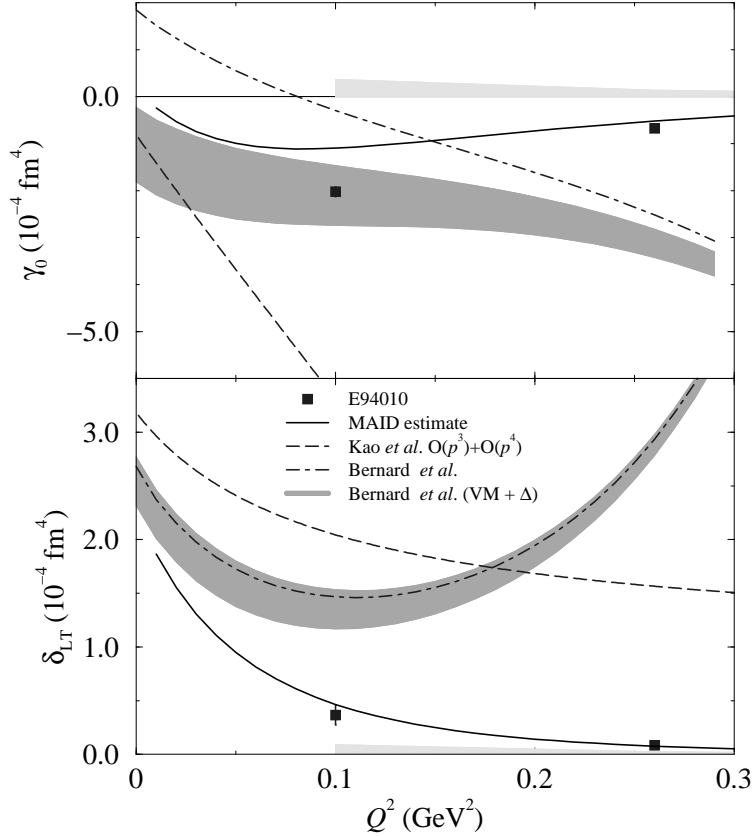}}}
\caption{Results for the neutron spin polarizabilities 
$\gamma _0$ (top-panel) and $\delta _{LT}$
(bottom-panel). Solid squares represent the results with statistical 
uncertainties. The light bands represent the systematic uncertainties. 
The dashed curves represent the heavy baryon$\chi $PT 
calculation~\protect\cite{chpt1}. 
The dot-dashed curves and the dark bands represent the relativistic baryon
$\chi $PT calculation 
without and with~\protect\cite{chpt2} the $\Delta $ and vector meson 
contributions, respectively.
Solid curves represent the MAID model~\protect\cite{dre01}.
}
\label{fig:fig3}
\end{figure}

The results for $\gamma_0(Q^2)$ are shown
in the top-left panel of Fig.~\ref{fig:fig3}. 
The statistical uncertainties are smaller than the
size of the symbols. 
The data are compared with 
a next-to-leading order ($O(p^4)$) HB$\chi$PT 
calculation~\cite{chpt1}, a next-to-leading order RB$\chi$PT
calculation and the same calculation explicitly including 
both the $\Delta$ resonance and vector meson contributions~\cite{chpt2}.
Predictions from the MAID model~\cite{dre01} are also shown.
At the lowest $Q^2$ point,
the RB$\chi$PT
calculation including the resonance contributions
is in good agreement with the experimental result.
For the HB$\chi$PT calculation without explicit resonance contributions, 
discrepancies are large even at $Q^2 = 0.1$ GeV$^2$. 
This might indicate the significance of the resonance contributions or a
problem with the heavy baryon approximation at this $Q^2$.
The higher $Q^2$ data point is in good agreement with the MAID
prediction,
but the lowest data point at $Q^2 = 0.1 $ GeV$^2$ is significantly lower.
Since $\delta_{LT}$ is 
insensitive to the
$\Delta$ resonance contribution, it was believed that $\delta_{LT}$ should be
more suitable than $\gamma_0$ to serve as a testing ground for the chiral 
dynamics of QCD~\cite{chpt1,chpt2}.
The bottom-left panel of Fig.~\ref{fig:fig3} shows $\delta_{LT}$ 
compared to
$\chi$PT calculations and the MAID predictions. While the MAID predictions are in good agreement with the results, it is surprising to see
that 
the data are in significant disagreement with the $\chi$PT calculations 
even at the lowest $Q^2$, 0.1 GeV$^2$. 
This disagreement (``$\delta_{LT}$ puzzle'') presents a significant challenge to the present Chiral Perturbation Theory.

Results of $gamma_0$ on the proton has been recently submitted for publication~\cite{eg1b}.  They show significant disagreement with both $\chi$PT 
calculations~\cite{chpt1,chpt2}. The isospin separation was performed and discussed in Ref.~\cite{eg1b}.

New experimental data have been taken at very low $Q^2$, down to 0.02 GeV$^2$
for the neutron ($^3$He)~\cite{e97110} for both longitudinal and transverse target polarizations, for the proton and deuteron~\cite{e03006} for only the longitudinal target polarization.
Preliminary results just became available for the neutron. 
Analysis is underway for the proton and deuteron data.
These results will shed light and provide benchmark tests to the
$\chi$PT calculations at the kinematics where they are expected to work.
A new proposal~\cite{e08027} was recently approved to measure $g_2^p$ with a transversely polarized proton target in the low $Q^2$ region. It will provide an isospin separation of the spin polarizabilities to shed light on the ``$\delta_{LT}$'' puzzle. 

\subsection{Precision measurements of $A_1$ in the high-$x$ region and 
polarized valence quark distribution}
\par
JLab Hall A experiment E99-117~\cite{e99117} measured $A_1^n$ with 
high precision in the $x$ region from 0.33 to 0.61
($Q^2$ from 2.7 to 4.8 GeV$^2$). 
Asymmetries from inclusive scattering of 
a highly polarized 5.7 GeV electron beam 
on a high pressure ($>10$ atm) (both longitudinally and
transversely) polarized $^3$He target were measured. 
Parallel and perpendicular asymmetries
were extracted for $^3$He. After taking into account the beam and target 
polarizations and the dilution factor,
they were combined to form $A_1^{^3He}$. Using the most recent 
model~\cite{model}, nuclear
corrections were applied to extract $A_1^n$. The results on $A_1^n$
are shown in the left panel of Fig. 5. 

The experiment greatly improved the precision
of data in the high-$x$ region, providing the first evidence that 
$A_1^n$ becomes positive at large $x$, showing clear SU(6) symmetry 
breaking. The results are in good agreement with the LSS 2001 pQCD
fit to previous world data~\cite{LSS2001} (solid curve) and 
the statistical model~\cite{stat} (long-dashed curve).
The trend of the data is consistent with the RCQM~\cite{vqm} predictions
(the shaded band). The data disagree with the predictions from the 
leading-order pQCD models~\cite{pQCD} (short-dashed and dash-dotted curves).
These data provide crucial input for the global fits to the world data to 
extract the 
polarized parton distribution functions and the extractions of 
higher-twist effects. 

\begin{figure}[!ht]
\parbox[t]{2in}{\scalebox{0.47}{\centering\includegraphics[10,-78][432,405]{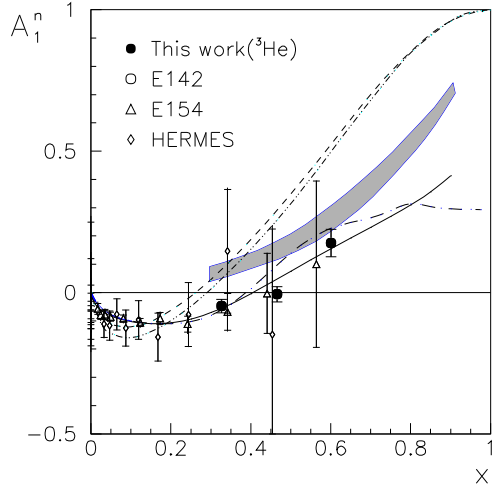}}}
\parbox[t]{2in}{\scalebox{0.47}{\centering\includegraphics[-20,-78][402,405]{dqoverq.epsi}}}
\vspace{-5mm}
\caption {A$_1^n$ (left-panel), $\Delta u/u$ (upper side of the right-panel) 
and $\Delta d/d$ (lower side of the right-panel) results from JLab Hall A 
E99-117~\protect\cite{e97110} and CLAS eg1b~\protect\cite{eg1b} experiments, compared with 
the world data, the JLab 12 GeV projections (open points)~\protect\cite{a1n12gev} and theoretical predictions~\protect\cite{feng07}.}
\end{figure}
\medskip

New results of $A_1^p$ and $A_1^d$ from the Hall B eg1b experiment~\cite{eg1b} 
have recently become available. The data cover the $Q^2$ range of 1.4 to 4.5 
GeV$^2$ for $x$ 
from 0.2 to 0.6 with an invariant mass larger than 2 GeV. 
The precision of the data
improved significantly over that of the existing world data.  

In the leading-order approximation,
the polarized quark distribution functions $\Delta u/u$ and $\Delta d/d$ were 
extracted from the Hall A neutron data, the CLAS eg1b proton and deuteron data
and the world data. 
The results are shown
in the right panel of Fig. 5, along with predictions from the 
leading-order pQCD (short-dashed curves) and pQCD fit including quark orbital angular momentum contributions~\cite{feng07}.  The results of $\Delta d/d$ are in 
significant disagreement with the predictions from a
leading-order pQCD model assuming hadron helicity conservation.
Data agree much better with the fit including quark-orbital angular momentum contributions, suggesting that the quark orbital
angular momentum may play an important role in this kinematic region.

\subsection{Precision $g_2$ measurements and higher twist effects}
\par
A precision measurement of $g_2^n$ from JLab Hall A E97-103~\cite{e97103} covered 
five different $Q^2$ values from 0.58 to 1.36 GeV$^2$ at x $\approx 0.2$. 
Results for $g_2^n$ are given in the left panel of 
Fig. 6. The light-shaded area in the plot 
gives the leading-twist contribution, obtained by fitting world data\cite{BB} and
evolving to the $Q^2$ values of this experiment. The systematic errors are 
shown as the dark-shaded area near the horizontal axis.   

\noindent
\begin{figure}[!ht]
\parbox[t]{2in}{\scalebox{0.28}{\rotatebox{-90}{\centering\includegraphics[730,30][1152,530]{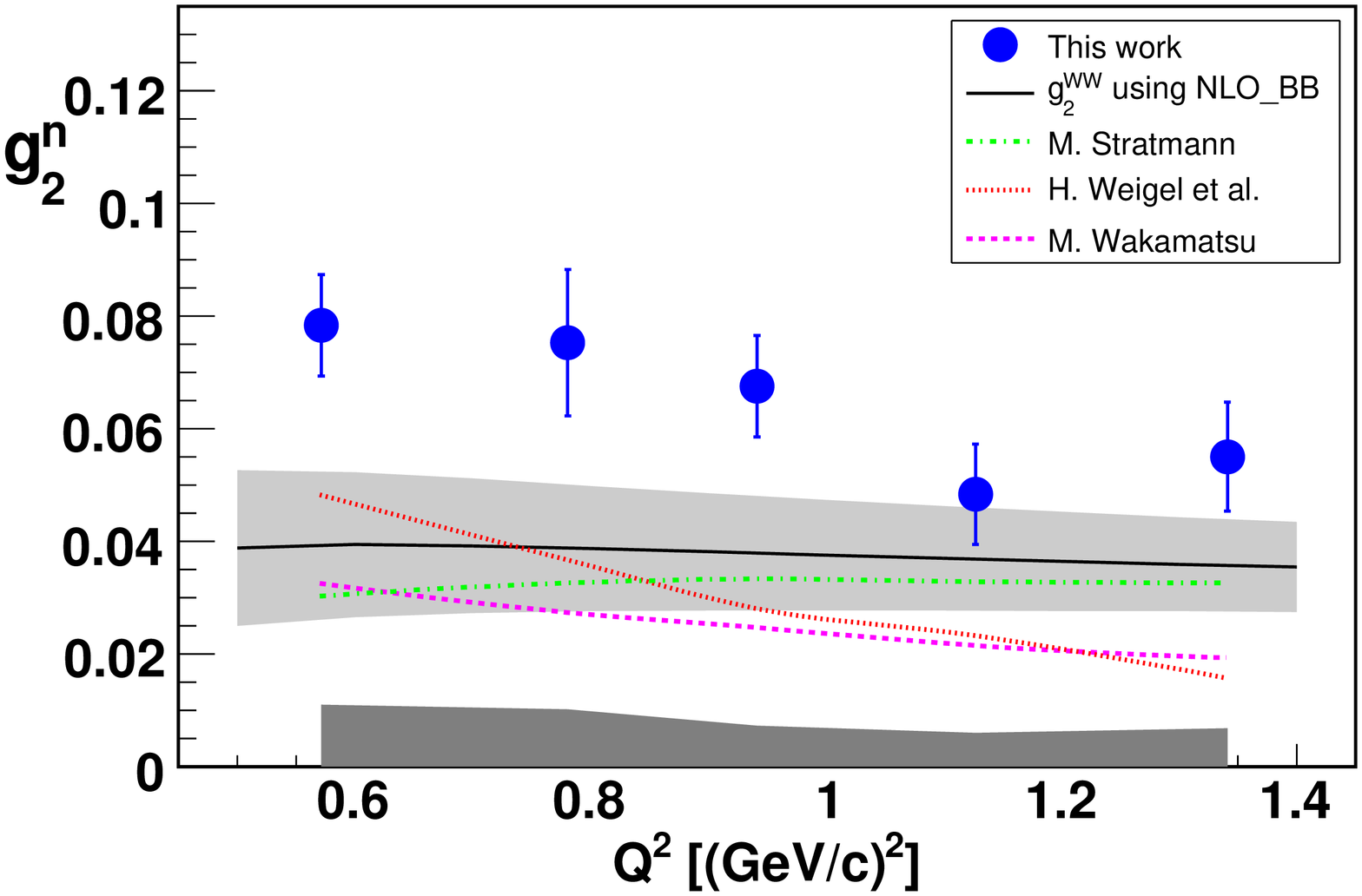}}}}
\parbox[t]{2in}{\scalebox{0.39}{\centering\includegraphics[-150,10][290,510]{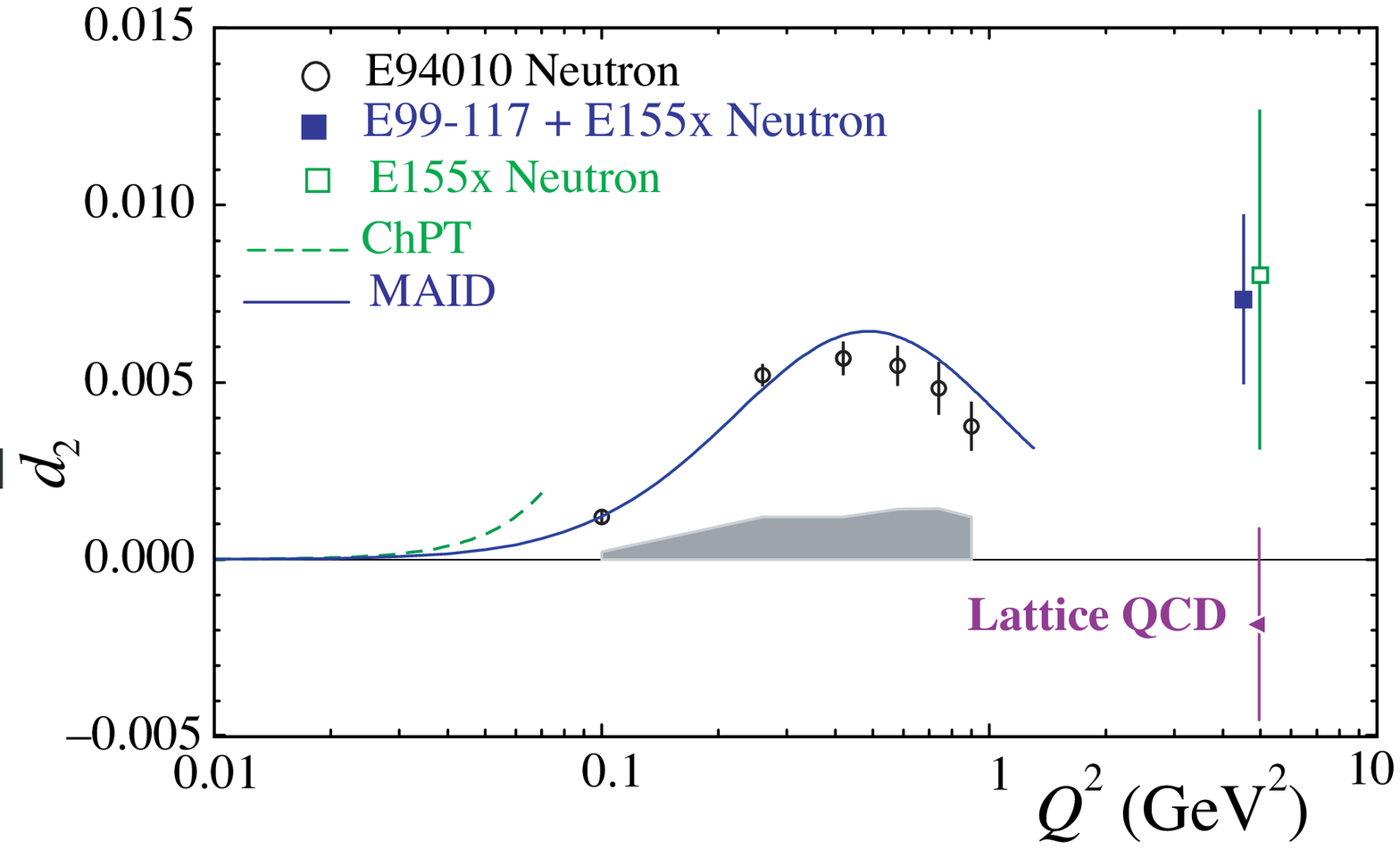}}}
\vspace {-50mm}
\caption{Results for $g_2^n$ (left panel) from JLab 
Hall A~\protect\cite{e97103}, in comparison with $g_2^{WW}$ and model 
predictions.
The right panel shows the $\bar d_2^n$ results from JLab Hall A~\protect\cite{e94010,e99117} and 
SLAC~\protect\cite{SLAC}, together with 
the Lattice QCD calculations~\protect\cite{LQCD}.}
\end{figure}
The precision reached is more than an order
of magnitude improvement over that of the best world data~\cite{E155x}.  The difference 
of $g_2$ from the leading twist part ($g_2^{WW}$)\cite{WW} is due to 
higher-twist effects and is sensitive to quark-gluon correlations. 
The measured g$_2^n$ values 
are consistently higher than g$_2^{WW}$.
For the first time, there is a clear indication that higher-twist effects 
become significantly positive at $Q^2$ below 1 GeV$^2$, 
while the bag model~\cite{str} and Chiral Soliton model~\cite{wei,wak} 
predictions of higher-twist effects are negative or close to zero. 
The $g_1^n$ data obtained from the same experiment agree with  
the leading-twist calculations within the uncertainties.

The second moment of the spin-structure function $d_2$ is of special interest: 
at high $Q^2$, it is a twist-3 matrix element and can be calculated in 
lattice QCD. Experimentally, due to $x^2$ weighting, the contributions are dominated by the high-$x$ region and the problem of low-$x$ extrapolation is avoided. 
The Hall A experiment E99-117 also provided data on $A_2^n$ at high-$x$. 
The precision of the $A_2^n$ data is comparable to that of the 
best existing world data~\cite{E155x} at high $x$. Combining these results with the world data, the second moment $d_2^n$ was extracted at an average $Q^2$ of 5
GeV$^2$.
Compared to the previously published result~\cite{E155x}, the uncertainty on $d_2^n$ has 
been improved by about a factor of 2. The $d_2$ moment at high $Q^2$ 
has been calculated by Lattice QCD\cite{LQCD} and a number
of theoretical models. While a negative or near-zero value was 
predicted by Lattice QCD and most models, the new result for $d_2^n$ 
is positive. Also shown in Fig. 2 are the low $Q^2$ (0.1-1 GeV$^2$) results of the inelastic part of $d_2^n$ from another Hall A experiment E94-010~\cite{e94010}, which were
compared to a Chiral Perturbation Theory calculation~\cite{chpt} and a model prediction\cite{maid}.

A Hall C experiment~\cite{RSS} measured $g_2$ on the proton and extracted $d_2^p$at a $Q^2$ value of 1.3 GeV$^2$. A more comprehensive measurement of $g_2^2$ and $d_2^p$ is planned in Hall C~\cite{SANE} later this year. It will cover a wide $Q^2$ range from 2.5 to 6.5 GeV$^2$.

\subsection{New results on spin-structure functions for quark-hadron duality study}
\par
JLab E01-012~\cite{e01012} ran successfully in early 2003 in Hall A. Asymmetries and cross sections were measured in the resonance region, in a $Q^2$ range from 1 to 3.6 GeV$^2$, for inclusive scattering of polarized electrons on a
longitudinally and transversely polarized $^3$He target. The spin-structure function $g_1$ and virtual photon asymmetry $A_1$ were extracted. The results for $A_1^{^3He}$ are presented in Fig. 7 (left panel). Also plotted are the world DIS data and a fit of the DIS data. It is interesting to note that the two sets of 
resonance data at the highest $Q^2$ agree well, indicating little or no $Q^2$-dependence, which is a key feature of the DIS data. These data also show the trend of becoming positive at the high-$x$ side, the same trend as observed for DIS data. The resonance data were integrated to study the global duality. Figure 5 (right panel) shows the results for both $^3$He and the neutron in comparison with the DIS fits evolved to the same $Q^2$. The resonance data agree with the DIS fits at least for $Q^2$ higher than 1.8 GeV$^2$, indicating that the global duality  holds for the neutron and $^3$He spin structure function, $g_1$, in the high Q$^2$ region.

\begin{figure}[!ht]
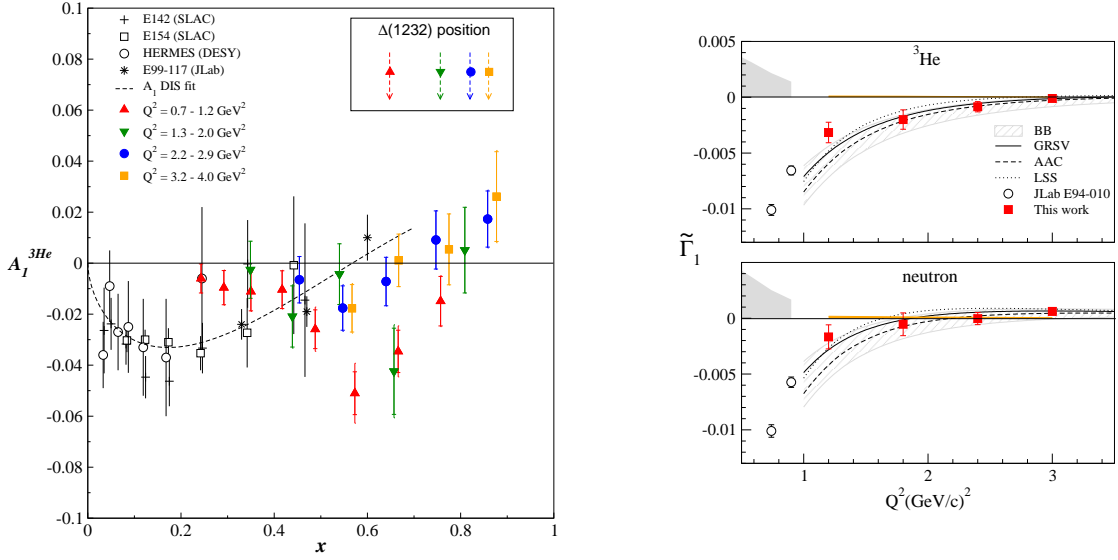

\parbox[t]{2in}{\scalebox{0.36}{\centering\includegraphics[30,50][452,533]{A1he3.eps}}}
\parbox[t]{2in}{\scalebox{0.40}{\centering\includegraphics[-230,122][192,605]{gamma1_duality.eps}}}
\vspace {5mm}
\caption {A$_1^{3He}$ (left panel) in the resonance region from JLab E01-012, compared with the world DIS data and a fit.
$\Gamma_1^{3He}$ and $\Gamma_1^n$ (right panel) of the resonance region from JLab E01-012, together with lower $Q^2$ results from JLab E94-010, compared with 
the world DIS fits.}
\vspace {-2mm}
\end{figure}

Results have also become available from the JLab Hall C experiment E01-006
\cite{RSS} and Hall B experiment eg1b~\cite{eg1b} on the 
proton and deuteron spin structure in the resonance region. These data, combined with the world DIS data, demonstrated that global quark-hadron duality holds 
in the proton and deuteron spin structure function at high $Q^2$ ($> 1.7$ GeV$^2$), while local duality seems violated in the $\Delta$ resonance region even for $Q^2$ values as high as 5 GeV$^2$.

\section{A planned measurement of neutron transversity at JLab}
\par
A recently approved JLab experiment~\cite{e06010} plans to measure the 
single-spin asymmetry of the ${\vec n}(e,e^\prime \pi^\pm)X$ reaction on 
a transversely polarized $^3$He target.
The goal of this experiment is to provide the first measurement 
of the neutron transversity, complementary to the HERMES and COMPASS 
measurements on the proton and deuteron.
This experiment focuses on the valence quark region, $x=0.19 - 0.34$,
at $Q^2=1.77 - 2.73$ GeV$^2$.  
Data from this experiment, when combined with data from 
HERMES~\cite{hermes03}, COMPASS~\cite{compass} and Belle~\cite{belle}, will provide powerful constraints on the transversity
distributions of both $u$-quarks and $d$-quarks in the valence region.

\begin{figure}
\centering{\scalebox{0.7}{\includegraphics{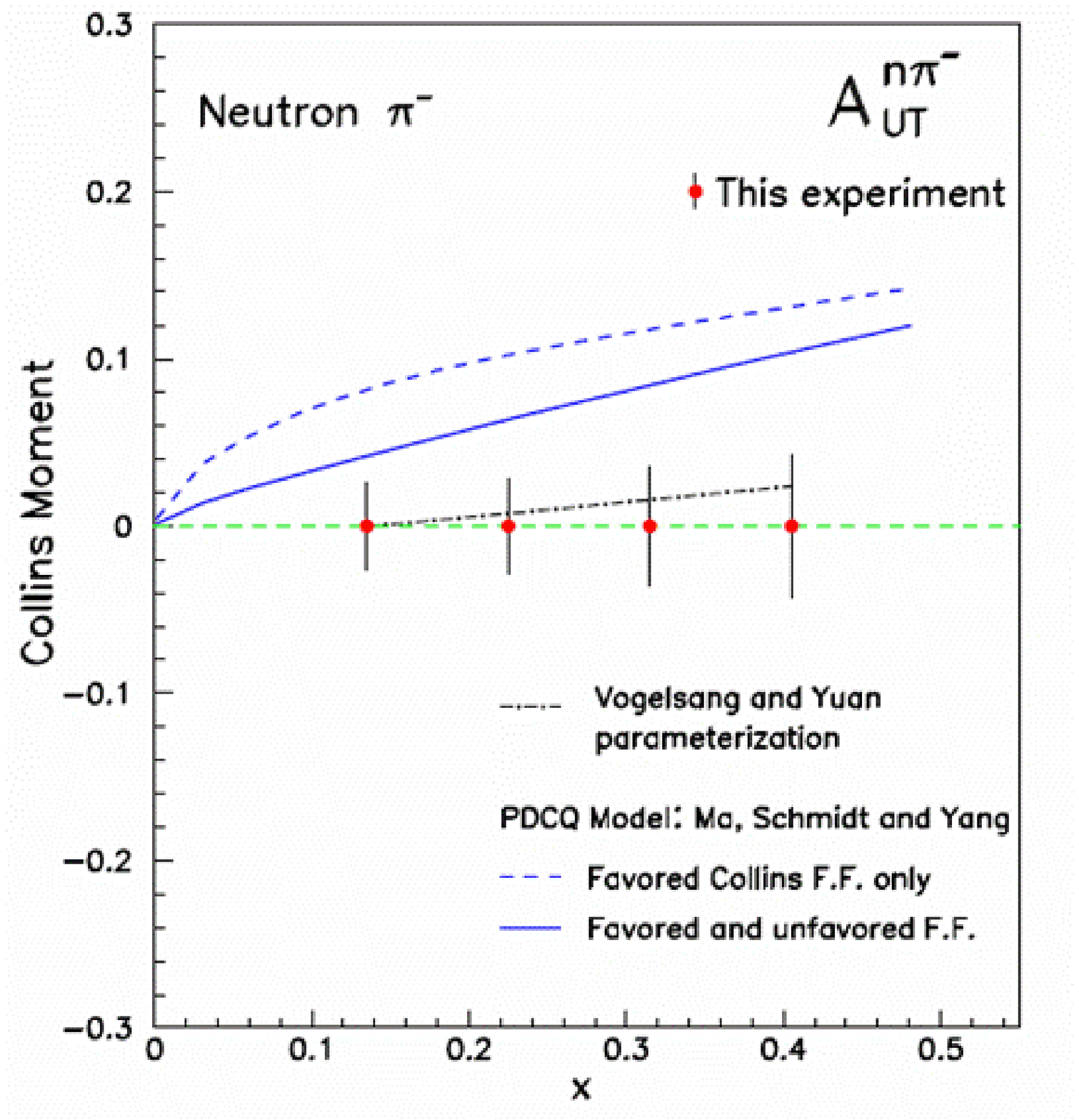}}}
\caption{Expected statistical precision of E06-010 for the Collins moment.} 
\label{fig:1}  
\end{figure}

The experiment will use a 6 GeV electron beam with the Hall 
A left-side high-resolution 
spectrometer (HRS$_L$) situated at 16$^\circ$ as the hadron arm,
and the BigBite spectrometer located at $30^{\circ}$ beam-right
as the electron arm.
A set of vertical coils will be added to the polarized $^3$He target 
to provide tunable polarization directions in all three dimensions. By
rotating the target polarization direction in the transverse plane, 
the coverage in $\phi_s^l$
is increased, hence facilitating the separation of the Collins
and the Sivers effects.
Figure~\ref{fig:1} shows the expected statistical precision of
this experiment with 29 days of beamtime for the ${\vec n}(e,e^\prime \pi^\pm)X$ 
single-spin asymmetry.
Due to the good particle identification in the HRS, $K^\pm$ data will be 
collected at the same time, providing a set of precision data to study
the transverse spin asymmetries for semi-inclusive $K^\pm$ production.

\section{Spin-structure program with the JLab 12 GeV energy upgrade}
\par

The JLab 12 GeV energy upgrade is the number 1 recommendation for future 
facilities in the US Nuclear Physics Long Range Plan published in 
December 2007~\cite{LRP07}. It has passed the US Department of Energy's 
Critical Decision 2 (CD2) review and is in the final stage of
engineering design. It is scheduled to start physics data taking in 2014.
The energy upgrade opens up a much wider DIS kinematics region to study 
nucleon spin structure. Planned experiments~\cite{a1n12gev} in the high-$x$ region
will definitively establish the contributions of valence quarks to the nucleon
structure. A precision measurement of the moment $d_2$~\cite{d212gev}, 
part of the 
the color polarizabilities, will provide a benchmark test of Lattice QCD 
predictions. An extensive SIDIS program~\cite{sidis} with transversely polarized neutron
and proton targets will map out precisely the Collins and Sivers moments.
Together with a world-wide effort, transversity distributions functions, 
Sivers distribution function and Collins-fragmentation functions will be
extracted. The tensor charge, a fundamental quantity of the nucleon, will be 
determined, which will provide a new benchmark test of Lattice QCD predictions.

A new facility, a Electron-Ion Collider (EIC), is undergoing discussion
in the US hadronic-physics community, as a long-term future facility. It will
provide unique capabilities for the study of QCD well beyond all existing 
facilities. It will extend the spin structure study 
over a very wide region.

\section{Summary}
\par
In summary, the high polarized luminosity available at
JLab has provided us with high-precision data to study the nucleon
spin structure in a wide kinematic range. They shed
light on the valence quark structure and help to 
understand quark-gluon correlations and study the non-perturbative region 
and the transition 
between perturbative and non-perturbative regions of QCD.
A planned precision study on transverse spin phenomena will open a new 
window to study the nucleon structure and help understand the
strong interaction.
\medskip

\footnotesize{
The work presented was supported in part 
by the U. S. Department of Energy (DOE)
contract DE-AC05-84ER40150 Modification NO. M175,
under which the
Southeastern Universities Research Association operates the 
Thomas Jefferson National Accelerator Facility.
}



\begin{thebibliography}{99}
\bibitem{spin} see, for example, B. W. Filippone and X. Ji, 
Adv. Nucl. Phys. {\bf 26}, 1 (2001)
\bibitem{Bjorken} J. D. Bjorken, Phys. Rev. {\bf 148}, 1467 (1966);
Phys. Rev. D{\bf 1}, 465 (1970)
\bibitem{e99117} X. Zheng, {\it et al.}, Phys. Rev. Lett. {\bf 92},
012004 (2004); X. Zheng, {\it et al.}, Phys. Rev. C {\bf 70}, 065207 (2004).
\bibitem{e97103} K. Kramer, {\it et al.}, Phys. Rev. Lett.{\bf 95},142002 (2005). 
\bibitem{chen05} J. P. Chen, A. Deur and Z. E. Meziani, Mod. Phys. Lett. A {\bf 20}, 2745 (2005).
\bibitem{e94010} M. Amarian, {\it et al.}, Phys. Rev. Lett. {\bf 89},
242301 (2002); 
{\it ibid.}, {\bf 92}, 022301 (2004);
{\it ibid.}, {\bf 93}, 152301 (2004); Z. E. Meziani {\it et al.}, Phys. Lett. B {\bf 613}, 148 (2005).
\bibitem{e94010he3} K. Slifer,{\it et al.}, submitted to Phys. Rev. Lett., arXiv:0803.2267 (2008).
\bibitem{bjsum} A. Deur, {\it et al.}, Phys. Rev. Lett. {\bf 93}, 212002 
(2004); A. Deur, {\it et al.}, submitted to Phy. Rev. Lett., arXiv:0802.3198 (2008).
\bibitem{e01012}P. Solvignon, {\em et al.},submitted to Phys. Rev. Lett., arXiv:0803.3845 (2008)
\bibitem{RSS} F. R. Wesselmann,  {\em et al.}, Phys. Rev. Lett. {\bf 98}, 132003 (2007).
\bibitem{SANE} JLab E07-003, S. Choi, Z. E. Meziani and O. Randon, spokesperosns. 
\bibitem{vqm} N. Isgur, Phys. Rev. D {\bf 59}, 034013 (1999).
\bibitem{pQCD}
S. Brodsky, M Burkhardt and I. Schmidt, Nucl. Phys. 
B{\bf 441}, 197 (1995).
\bibitem{SU6} F. Close, Nucl. Phys. B {\bf 80}, 269 (1974).
\bibitem{WW}
S. Wandzura and F. Wilczek, Phys. Lett. B 72 (1977).
\bibitem{d2}
X. Ji and W. Melnitchouk, Phys. Rev. D {\bf 56}, 1 (1997).
\bibitem{gdh} S. B. Gerasimov, Sov. J. Nucl. Phys. {\bf 2}, 598 (1965);
S. D. Drell and A. C. Hearn, Phys. Rev. Lett. {\bf 162}, 1520 (1966).
\bibitem{dre01} D. Drechsel, S.S. Kamalov, and L. Tiator, Phys. Rev. D {\bf 63}, 114010 (2001). 
\bibitem{ggdh} X. Ji and J. Osborne, J. Phys. G {\bf 27}, 127 (2001).
\bibitem{BG} E. D. Bloom and F. J. Gilman, Phys. Rev. Lett. {\bf 25}, 1140 (1970).
\bibitem{F2dual} I. Niculescu, {\it et al.}, Phys. Rev. Lett. {\bf 85}, 1182 (2000); {\bf 85}, 1186 (2000).
\bibitem{MEK} W. Melnitchouk, R. Ent and C. Keppel, Phys. Rept. {\bf 406}, 127 (2005).
\bibitem{HERMES} A. Airapetian, {\it et al.}, Phys. Rev. Lett. {\bf 90}, 092002 (2003); Eur. Phys. J. C {\bf 26}, 527 (2003); Phys. Rev. D {\bf 75}, 012007 (2007).
\bibitem{NIMA}Hall A collaboration: J. Alcorn \emph{et al.}, 
Nucl. Inst. Meth. A \textbf{522}, 294 (2004).
\bibitem{NIMB}
CLAS collaboration: B. A. Mecking \emph{et al.},
Nucl. Inst. Meth. A \textbf{503}, 513 (2003).
\bibitem{NH3} 
C. D. Keith \emph{et al}., Nucl. Inst. Meth. A \textbf{501}, 327 (2003).
\bibitem{model} F. Bissey, {\it et al.}, Phys. Rev. C {\bf 65}, 064317 (2002).
\bibitem{LSS2001} E. Leader, A. V. Sidorov and D. B. Stamenov, Eur. Phys. J.
C {\bf 23}, 479 (2002).
\bibitem{stat} C. Bourrely, J. Soffer and F. Buccella, Eur. Phys. J. 
C {\bf 23}, 487 (2002).
\bibitem{feng07} H. Avakian, S. Brodsky, A. Deur, and Y. Feng, arXiv:0705.1553 (2007). 
\bibitem{E155x} K. Abe, {\it et al.}, E155 collaboration, Phys. Lett.
B {\bf 493}, 19 (2000).
\bibitem{BB} J. Bl{\"{u}}mlein and H. B\"{o}ttcher, Nucl. Phys. B {\bf 636}, 225 (2002).
\bibitem{str} M. Stratmann, Z. Phys. C {\bf 60}, 763 (1993).
\bibitem{wei} H. Weigel, Pramana {\bf 61}, 921 (2003).
\bibitem{wak} M. Wakamatsu, Phys. Lett. B {\bf 487}, 118 (2000).
\bibitem{LQCD} M. G\"{o}ckeler {\it et al.}, Phys. Rev. D {\bf 63}, 074506 (2001).
\bibitem{chpt}X. Ji, C. Kao, and J. Osborne, Phys. Lett. B {\bf 472}, 1 (2000).
\bibitem{chpt1}C.~W.~Kao, T. Spitzenberg and M. Vanderhaeghen, Phys. Rev. D {\bf 67}, 016001 (2003).
\bibitem{chpt2}V. Bernard, T. Hemmert and Ulf-G. Meissner, Phys. Rev. D {\bf 67}, 076008 (2003).
\bibitem{maid} D. Drechsel, S. Kamalov and L. Tiator, Phys. Rev.  D {\bf 63}, 114010 (2001)
\bibitem{ciofi} C. Ciofi degli Atti and S. Scopetta, Nucl. Phys. B{\bf 404}, 
223 (1997)
\bibitem{ciofi2} C. Ciofi degli Atti, E. Pace and G. Salme, Phys. Rev. C{\bf 51}, 1108 (1995).  
\bibitem{TB} E. Thomas and N. Bianchi, Nucl. Phys. B{\bf 82} (Proc. Suppl.), 
256 (2000).
\bibitem{BC} H. Burkhardt and W. N. Cottingham, Ann. Phys. {\bf 56}, 453 (1970)
\bibitem{alphas} A. Deur, V. Burkert, J. P. Chen and W. Korsch, Phys. Lett. B
{\bf 650},4 244 (2007); submitted to Phys. Lett. B, arXiv:0803.4119 (2008). 
\bibitem{ads} S. J. Brodsky and G. F. de Teramond, Phys. Rev. Lett. {\bf 96}, 201601 (2006).
\bibitem{e97110} JLab E97-110, Spokespersons, J. P. Chen, A. Deur and 
F. Garibaldi.
\bibitem{e03006} JLab E03-006, Spokespersons, M. Battaglieri, A. Deur, R. De Vita and M. Ripani.  
\bibitem{e08027} JLab E08-027, Spokespersones, A. Camsonne, J. P. Chen and K. Slifer.
\bibitem{eg1a} R. Fatemi {\em et al.}, Phys. Rev. Lett. {\bf 91}, 222002 (2003); J. Yun {\em et al.}, Phys. Rev. {\bf C 67}, 055204 (2003). 
\bibitem{eg1b} K.V. Dharmawardane {\em et al.}, Phys. Lett. {\bf B 641} 11 
(2006); Y. Prok {\em et al.}, submitted to Phys. Rev. Lett, arXiv:0802.2232 (2008); P. Bosted, {\em et al.}, Phys. Rev. {\bf C 75}, 035203, (2007).
\bibitem{dre} D. Drechsel, B. Pasquini and M. Vanderhaeghen, Phys. Rep. 
{\bf 378}, 99 (2003); D. Drechsel and L. Tiator, Ann. Rev. Nucl. Part. Sci.
{\bf 54}, 69 (2004).
\bibitem{JU} X. Ji and P. Unrau, Phys. Lett. {\bf B 333}, 228 (1994).
\bibitem{SLAC} K. Abe {\em et al.}, Phys. Rev. {\bf D 58},112003 (1998); 
P. L. Anthony, {\em et al.}, Phys. Lett. {\bf B 493}, 19 (2000), {\bf B 553}, 18 (2003).
\bibitem{F2&RJLab}
Y. Liang et al., nucl-ex/0410027.
\bibitem{sof02}J. Soffer and O. V. Teryaev, Phys. Rev. {\bf D 70}, 116004 (2004).
\bibitem{bur92}V. D. Burkert and B. L. Ioffe, Phys. Lett. {\bf B 296}, 223 (1992).
\bibitem{d26gev} JLab experiment E06-014, S. Choi, X. Jiang, Z.E. Meziani and B. Sawatzky spokespersons.
\bibitem{BST}C. Bourrely, J. Soffer and O. V. Teryaev, Phys. Lett. {\bf B420}, 375 (1998).
\bibitem{soffer} J. Soffer, Phys. Rev. Lett. {\bf 74}, 1292 (1995). 
\bibitem{compass} The COMPASS collaboration, Phys. Rev. Lett. {\bf 94}, 202002 (2005); Nucl. Phys. {\bf B765}, 31-70 (2007). 
\bibitem{mulders96} P. Mulders and R. D. Tangerman, Nucl. Phys. {\bf B461}, 197 (1996).
\bibitem{collins} J. Collins, Nucl. Phys. {\bf B396}, 161 (1993).
\bibitem{sivers} D. W. Sivers, Phys. Rev. {\bf D41}, 83 (1990).
\bibitem{hermes03} A. Airapetian, 
{\it et al.}, Phys. Rev. Lett. {\bf 94}, 012002 (2005).
\bibitem{belle} A. Ogawa, {\it et al.}, Proceedings of DIS2005, AIP {\bf 792}, 949 (2005).
\bibitem{e06010} JLab E06-010, J. P. Chen, H. Gao, E. Cisbani, X. Jiang, J.-C. Peng, Spokjespersons.
\bibitem{LRP07} The Nuclear Science Advisory Committee Report: ``The Frontiers of Nuclear Science - A Long Range Plan'', (2007)
\bibitem{d212gev} JLab experiment E12-06-121, T. Averett, W. Korsch, Z.E. Meziani and B. Sawatzky, spokespersons.
\bibitem{a1n12gev} JLab experiment E12-06-110, 
G. Cates, J.P. Chen, Z.E. Meziani and X. Zheng, spokespersons; JLab E12-06-109,
S. Kuhn, {\it et al.}, spokespersons.
\bibitem{sidis} A. Afanasev, {\it et al.}, arXiv:hep-ph/0703288 (2007).

\end{thebibliography}
\end{document}